\documentclass[screen]{acmart}
\AtBeginDocument{%
  \providecommand\BibTeX{{%
    \normalfont B\kern-0.5em{\scshape i\kern-0.25em b}\kern-0.8em\TeX}}}

\begin{document}

\title{The Influence of Color Stimuli on Adolescents' Emotion Playing Mobile Games}

\author{Leonie Kallabis}
\email{leonie.kallabis@th-koeln.de}
\orcid{0000-0001-7131-657X}
\affiliation{%
  \institution{TH Köln}
  \streetaddress{Steinmüllerallee 6}
  \city{Gummersbach}
  \country{Germany}
  \postcode{51643}
}

\author{Bruno Baruque-Zanón}
\email{bbaruque@ubu.es}
\orcid{0000-0002-4993-204X}
\affiliation{%
  \institution{University of Burgos}
  \streetaddress{Avda. Cantabria s/n}
  \city{Burgos}
  \country{Spain}
  \postcode{09006}
}

\author{Heinrich Klocke}
\email{heinrich.klocke@th-koeln.de}
\orcid{0000-0002-4204-7371}
\affiliation{%
  \institution{TH Köln}
  \streetaddress{Steinmuellerallee 6}
  \city{Gummersbach}
  \country{Germany}
  \postcode{51643}
}

\author{Ana María Lara-Palma}
\email{amlara@ubu.es}
\orcid{0000-0002-0127-7963}
\affiliation{%
  \institution{University of Burgos}
  \streetaddress{Avda. Cantabria s/n}
  \city{Burgos}
  \country{Spain}
  \postcode{09006}
}

\author{Boris Naujoks}
\email{boris.naujoks@th-koeln.de}
\orcid{0000-0002-8969-4795}
\affiliation{%
  \institution{TH Köln}
  \streetaddress{Steinmüllerallee 6}
  \city{Gummersbach}
  \country{Germany}
  \postcode{51643}
}

\renewcommand{\shortauthors}{Kallabis, et al.}

\begin{abstract}
Video games elicit emotions which can be influenced by color stimuli as shown by previous studies. However, little research has been conducted on whether this applies to mobile games played by adolescents.
Therefore, we examined the influence of color stimuli hue and saturation on mobile game play. Adolescents (n=21) played a mobile platformer game with varying hue and saturation per level for about 25 minutes. We gathered data on emotional states after each level using the Self-Assessment Manikin questionnaire, recorded time spent in each level, and collected participant self-reports on their video game experience. 
We performed statistical tests, such as ANOVA, which depict no significant influence of hue and/or saturation on the emotional state of our players. We conclude that it is possible that color alone is not an effective measure for eliciting emotion in mobile games, and further research is needed to consider measures such as time spent in the game and screen size, as these are unique to mobile games.
There was a noticeable variance in emotional response between male and female players, with a significant interaction of hue and saturation among male players for valence ratings. This may be an indication that color preference influences perceived pleasantness.

\end{abstract}

\keywords{Games/Play, Teens, Mobile Devices: Phones/Tablets, Empirical study that tells us about people, Emotion, Color Stimuli}

\maketitle
\pagestyle{plain}

\section{Introduction}
\label{sec:intro}

How can we use color stimuli to influence players' emotions in mobile games? 
Over the years, the emotional experiences that video games are designed to create have evolved. We now have games that require moral choices, that aim to elicit emotional responses in general, or that target discrete emotions such as fear in horror games~\cite{Hemenover2018}. 
Games consist of numerous elements that affect players' emotions, including narrative~\cite{Frome2007}, color~\cite{Joosten2012, Geslin2016, Plass2020}, lighting~\cite{Knez2008}, game mechanics~\cite{Pawar2019}, sound~\cite{Pawar2019}, and more. 
In psychology, the relationship between color and emotion has been a focus of research since the 18th century~\cite{Elliot2019}. There is a considerable amount of research investigating the psychological arousal of using color in different contexts. Findings include links between specific colors and emotions, such as red and saturation or chroma leading to arousal or stimulation~\cite{Elliot2019}. 

Research on color in video games and its relationship to gender and age is sparse, but the relationship between gender, age, and video games, and gender, age, and color has been fairly well researched. Color preference research shows that it is possible to find generalizable results, at least in Western countries. Looking at gender and age, these are mixed, with some finding men's preferences for blue and women's for red, but others finding similarities between genders~\cite{Elliot2019}. In the gaming context, gender and age differences were found in terms of time spent playing, motivation to play, and genre preference.~\cite{Greenberg2010,Tondello2019}. 

Previous studies that have examined the impact of color on emotion in video games have found that different emotions can be elicited~\cite{Joosten2012, Geslin2016, Plass2020}. Additionally, the use of emotions is gaining attention in multimedia learning and, more recently, in games for learning and serious games, especially through emotional design. Emotional design is using a range of design features to impact learners' emotions and enhance learning outcomes~\cite{Plass2016, Pawar2019}. The use of color in previous research has shown promising results in terms of emotion elicitation and learning outcomes. For example, the use of pleasant colors in combination with anthropomorphism~\cite{Brom2018}, or the use of warm colors to induce happy emotions and neutral colors to induce sad emotions in adolescents~\cite{Plass2020}.

Video games are available in many different forms on varying devices. Since the advent of smartphones, the mobile game market has grown significantly. Understandably so, as smartphones make games available anytime and anywhere. Additionally, they make it possible for a wide range of young people to access games. Mobile games are also developed in the areas of serious games~\cite{Ouariachi2018} and game-based learning~\cite{Chang2019}. These offer opportunities to provide learning experiences or opportunities without being tied to time and space. Previous studies~\cite{Joosten2012, Geslin2016, Plass2020, Wilms2018} have focused on the effect of color on larger screens, or surfaces, relative to those of conventional smartphones. In contrast, mobile games are used in very different places. Considering the Color-in-Context Theory~\cite{Elliot2012}, the effect of color could be different based on the before-mentioned circumstances.
In addition, by looking at multiple dimensions of color: hue and saturation, we aim to extend previous research on color in games, that has focused only on hue~\cite{Joosten2012, Plass2020}.

Understanding how color influences adolescents' emotions in the context of mobile games can improve our understanding of the use of color in game design. For example, in learning games, where we could improve information representation~\cite{Plass2016}, motivation, retention, comprehension, and transfer~\cite{Brom2018}, as well as increase overall positive emotions.
This paper aims to provide an overview of the current literature on color research and color research in game design, educational game design, and mobile games in particular. We conducted an initial study to provide initial findings on the use of color in mobile game experiences and to provide suggestions for further research in this area.

In the following Section~\ref{sec:background}, we review additional research related to emotions, color, video games, and mobile games. We then derive our objectives and research questions in Section~\ref{sec:aims_rqs}. This is followed by a description of the prototype (Section~\ref{sec:prototype}) and the methodology (Section~\ref{sec:methodology}). We then describe the results of the study (Section~\ref{sec:results}), followed by a discussion and analysis of the main findings (Section~\ref{sec:analysis}). We conclude with a summary and opportunities for future research in Section~\ref{sec:conclusion}.


\section{Background}
\label{sec:background}
The following section provides background literature on emotion, color, the relationship between emotion and color, and video games, mobile games, and emotion. It also considers the importance of age, gender, and video game experience in this context and serves as a baseline for the initial study conducted.

\subsection{Emotion}
It is important to distinguish between the terms emotion and mood in order to understand which concept this paper is primarily concerned with. According to Izard, emotion involves neural circuits, response systems, and a feeling state or process that motivates and organizes cognition and action, while also being typically related to an object, such as being afraid of something~\cite{Izard2010}. Mood, on the other hand, reflects the general state of an individual's situation. This goes hand in hand with the fact that emotions tend to be short-lived, whereas a mood can last for several hours or days~\cite{Hemenover2018}. 

Emotions are conceptualized in different models. Ekman identified six basic emotions, namely anger, fear, disgust, happiness, sadness, and suprise~\cite{Ekman2004}. Each of these emotions describe a single feeling state with a biological mechanism.
Another concept to describe emotion is per dimensions. Russell conceptualized emotion based on the two dimensions: arousal and valence~\cite{Russell1980}. He created the circumplex model of affect that can be used to classify emotion based on the expressions perceived by a person for arousal and valence. Arousal can be seen as being activated in response to a situation. A high arousal exists when people are in a proactive status, a low one when they are in a reactive status. Valence is a feeling of pleasure and displeasure that is expressed in response to an external event and extends accordingly between these areas. 

\begin{figure}[ht]
    \centering
    \includegraphics[width=0.45\textwidth]{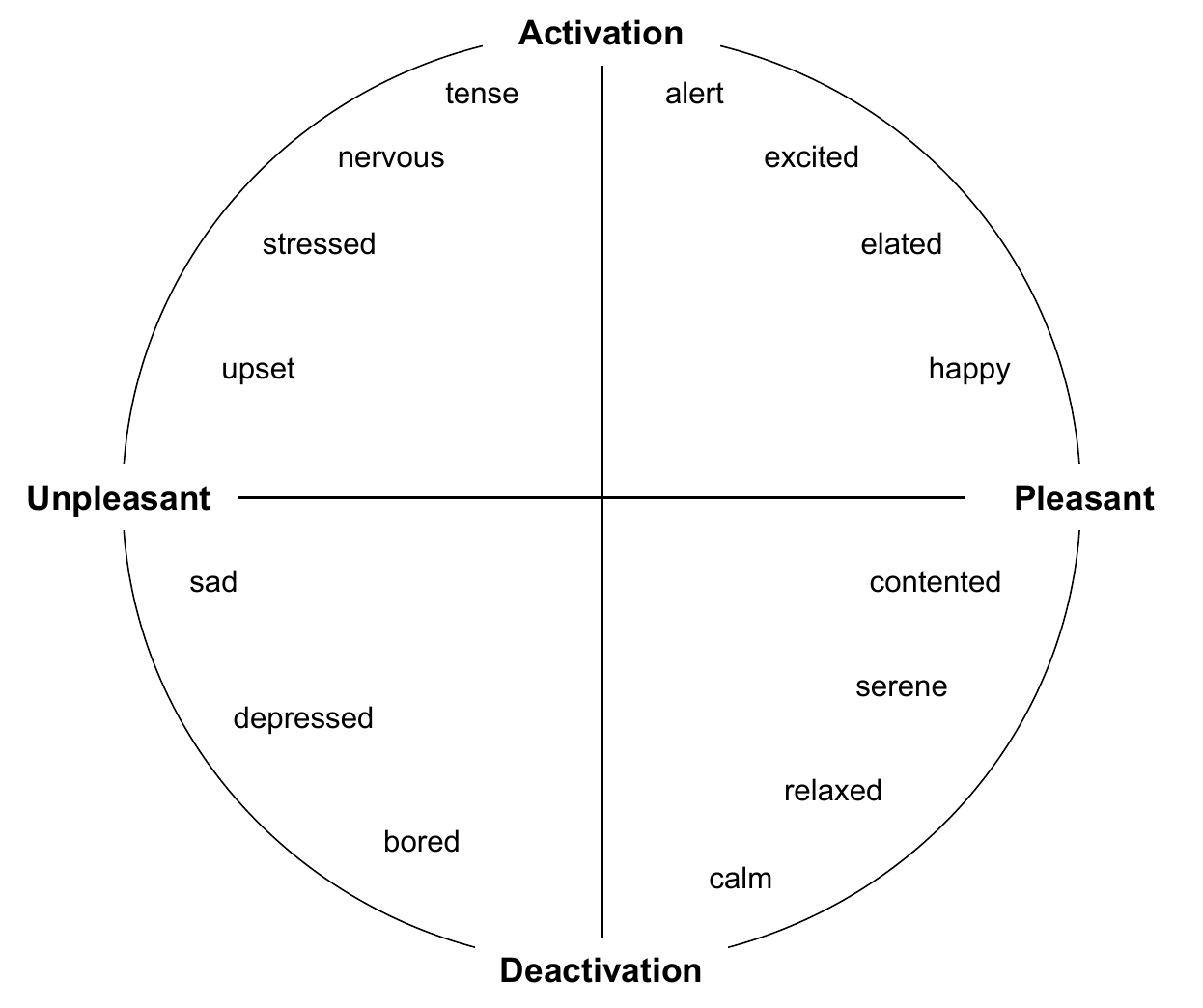}
    \caption{Recreated graphical representation of the Circumplex model of affect as depicted in~\cite{Posner2005}. The horizontal axis represents the valence dimension, the vertical axis represents the arousal dimension.}
    \Description{Recreated graphical representation of the Circumplex model of affect as depicted in~\cite{Posner2005}. The horizontal axis represents the valence dimension, the vertical axis represents the arousal dimension.}
    \label{fig:circumplex_model}
\end{figure}

The determination of an affective state can be achieved by combining the values of arousal and valence using the circumplex model, as shown in figure~\ref{fig:circumplex_model}. For example, a little over medium arousal with high valence is the representation for a happy emotion. 

\subsection{Color and Emotion}

The dimensions of color, theories of color and emotion, and the importance of culture in relation to color are presented below.

\paragraph{Color Dimensions}
Color has multiple properties, which are often called dimensions in this context. Three of these dimensions are hue, brightness, and saturation, where brightness has similar features to another dimension called lightness and saturation has similar features to a dimension called chroma~\cite{Wyszecki2000, Fairchild2015}. 
In the following, the term color is used interchangeably for the combination of the dimensions hue, saturation and brightness. Hue refers to the characteristic generally associated with a color, such as red, green, or blue. Saturation describes the difference from an achromatic stimulus; for example colors with low saturation have only a slight difference from the corresponding gray color. Brightness refers to the perceived intensity of light~\cite{Wyszecki2000}.
When studying color, it is crucial to account for all three dimensions as research suggests that saturation and brightness have a significant impact on emotional response, potentially even greater than that of hue~\cite{Wilms2018}.

Wilms and Oberfeld presented combinations of hue, saturation, and brightness on an LED display to 62 participants. The environment was controlled for light, acoustics, and electric signals. Emotional response was measured using skin conductance, heart rate, and the Self-Assessment Manikin (SAM) questionnaire as a self-report measure~\cite{Wilms2018}. The study found that a combination of saturated and bright colors led to higher arousal, with the highest arousal perceived for red compared to green and blue. Valence ratings were highest when highly saturated and bright colors were displayed.
Interaction effects between color properties were also observed, with highly saturated blue receiving the highest valence ratings and low saturated blue receiving the lowest valence ratings. The identified interactions highlight the significance of the connections between hue, saturation, and brightness.

\paragraph{Color-in-Context Theory}
The Color-in-Context Theory is a valuable tool for examining the link between color and emotion, providing six premises that outline the psychological impact of color\cite{Elliot2012}: 
\begin{enumerate}
\item \textit{Color carries meaning:} describes that color can have a functional as well as an aesthetic value
\item \textit{Viewing color influences psychological functioning}
\item \textit{Color effects are automatic:} color information is processed without intention or awareness
\item \textit{Color meanings (and associated responses) have two sources: Learning and biology}: We "learn" color preferences while biological perceptions reinforced survival
\item \textit{Relations between color perception and affect, cognition and behavior are reciprocal}
\item \textit{Color meanings and effects are context specific}
\end{enumerate}

\paragraph{Color and Affect}
Relating to premises (2) and (3) Elliot reports on extensive research exploring the psychological arousal induced by color in various contexts as printed paper, light projected in or into a room, a painted wall, a computer screen, or using semantic color words. Research has revealed connections between certain colors and emotions, such as red and saturation or chroma leading to arousal or stimulation, and blue and green leading to calmness or relaxation. Links between yellow or orange, brightness, and excitement or stimulation were found, but currently there is little support for these links~\cite{Elliot2019}. 

\paragraph{Color Preference}
Color preference (relating to premise (4)) is a topic that has been moderately researched, with a focus on adults and children. The question of whether this can be systematically described arose before the 20th century, arguing that it is a product of individual differences. However, consistent patterns were found in Western samples indicating that blue or blueish colors were the most preferred, yellow or yellowish were the least preferred, and green or greenish and red or reddish colors were in the middle~\cite{Elliot2019}. 

\paragraph{Cultural Meaning of Color}
Color preferences have been influenced by culture since antiquity. With respect to premise (4), research attributes reactions to colors to be either learned from personal experience or induced by biological descents~\cite{Aslam2006}. 
It is important to consider both views and to keep in mind that culture (i.e. values, language, religion, age, referents, gender and ethnicity) could play a role in people's color perception and affective response. For example, the color red has different associations in Anglo-Saxon (masculine, love, lust, fear, anger), Germanic and Slavic (fear, anger, jealousy), Latin (masculine), Nordic (positive), Chinese (love, happiness, lucky), Japanese (love, anger, jealousy) and Korean (love, adventure, good taste) countries~\cite{Aslam2006}.


\subsection{Emotion and Video Games}

Video games may elicit distinct emotional responses, such as high levels of fear among players engaging with horror games. Moreover, Hemenover and Bowman provide insights into literature exploring guilt and regret, as well as aggression and frustration, as discrete emotions that have been subject to research~\cite{Hemenover2018}.
From multimedia learning, different models of using emotions in learning and later in game-based learning emerged.
In relation to game-based learning (GBL), the Emotional Foundations of GBL (EmoGBL) was created, which aims to account for the emotive diversity of a game experience, while highlighting common mechanisms of these emotions that can guide the design of emotional gaming experiences~\cite{Loderer2020}. The authors provide implications for the emotional design of GBL environments, identifying five aspects: visual aesthetic design, musical score, game mechanics, narrative, and incentive system. Detailing visual aesthetic design, the authors elaborate on some fundamental emotion-relevant features, such as shape and color. 
Game characters or virtual agents that serve as instructors or companions in educational games have also been a focus of research. Here, the visual appearance has been analyzed, for example, physical attractiveness or realism, as reported by~\cite{Loderer2020}. The effect of different colors on characters has also been studied~\cite{Plass2020}.


\subsection{Age, gender and video game experience}
Given the recognized differences in cultural perceptions of color as well as varying preferences for video games according to age and gender, it is important to investigate possible color effects in video games.
%
Studies looking at gender and age differences regarding color show mixed results, especially for age there is not much research available. In their review Elliot report that some literature found that men prefer blue and women prefer red in Western countries, but other literature has found similarities across genders~\cite{Elliot2019}.

Greenberg et al. found that age and gender have an impact on game playtime, motivation to play and genre preference. They suggest that this may be due to children and adolescents’ age-dependent differing development stages~\cite{Greenberg2010}. Recently, Tondello and Nacke identified a difference in game genre preferences based on age and gender. Their findings suggest that men are more drawn to games that are challenging and competitive, while women are more attracted towards games that are immersive and relaxing~\cite{Tondello2019}.
Joosten et al. discovered that novice players exhibit a greater emotional response to changes in color within the gaming environment, as opposed to experienced players. This can be explained by the fact that inexperienced players tend to depend more heavily on visual cues for guidance, resulting in a greater degree of attention being paid to the visual aspects of the game. Another possible explanation is that experienced players spent less time playing each color condition, and therefore may have had fewer emotional responses. ~\cite{Joosten2012}. 

\subsection{Mobile Games}
\label{subsec:mobile_games}
Mobile games are not limited to playing on a mobile device, but also include games that are portable like card games and can be played in different contexts. However, with the advent of the smartphone, the mobile gaming experience on mobile phones has evolved rapidly~\cite{Kowert2020}. Accordingly, we will use the term mobile game to refer to games played on mobile phones.
Mobile phones have technological limitations such as smaller screens, limited input capabilities, and lower hardware performance. These limitations lend themselves to casual games and have made mobile games successful and reach a wider audience~\cite{Kowert2020, Anable2018}. Casual games have appealing content, simple controls, quick rewards, and an easy to learn gameplay~\cite{Kuittinen2007}.
Compared to other games, mobile gaming sessions require no preparation, making them playable anytime, anywhere. They are also designed to be played in short time frames like five to ten minutes~\cite{Anable2018}.


\subsection{Understanding Color and Emotion in Video Games}
Just a few studies researched the effect of color on emotions in video games.
Geslin et al. examined the emotional influence of 24 video game images among a sample of 85 individuals~\cite{Geslin2016}. They used a subjective semantic questionnaire to evaluate perceived emotions. The study found a meaningful correlation among luminance, saturation, and lightness and emotions of joy, sadness, fear, and serenity. Greater saturation elicited a higher valence, while less saturated images resulted in negative valence and instilled fear. Low brightness is more likely to elicit fear in players, while higher brightness inspires confidence. Moreover, a combination of diverse colors can evoke positive feelings of joy. Although the findings are interesting, they pertain to reactions to images and not a dynamic gaming experience. Therefore, further investigation is required to validate their relevance in a game experience.

Plass et al. investigated how color, shape, expression, and dimensionality of game characters elicit emotion in digital games for learning~\cite{Plass2020}. The emotional response of participants was measured via self-reports. The results indicated a medium effect size for color on perceptions of affective quality. Warm colors (such as orange) elicited positive emotions, and neutral colors (such as gray) elicited negative emotions. The results regarding color are the outcome of three forced choice paradigm studies in which participants compared images of video game characters. As for the study of Geslin et al. it would be interesting to see if those effects also hold in a gaming experience.

Joosten et al. designed a game in which the ambient light color can be varied to elicit a specific emotion. To assess the elicited emotions of 51 participants, the Self-Assessment Manikin (SAM) questionnaire was employed~\cite{Joosten2012}. They found that the color red triggers high arousal in combination with low valence. They also reported that yellow induced high valence ratings in the participants. Both results were significantly different from the emotional responses they measured for other colors.

\section{Aims and Research Questions}
\label{sec:aims_rqs}
Plass et al. and Joosten et al. examined hue, but evidence suggests that saturation also significantly affects emotion~\cite{Elliot2019}. We aim to expand upon this research by examining not only hue but also saturation.

Emotions in video games have been a focus of both academic research and the game industry. Understanding the production of discrete emotions is an important aspect in this area. The complex nature of video games implies that several factors can influence their emotional impact. Therefore, scrutinizing specific design characteristics and evaluating their potential to evoke emotions could be beneficial, as currently explored through emotional design~\cite{Plass2020,Loderer2020}. As mobile games grow in popularity and differ from games played on computers or consoles in terms of screen size, input capabilities, and hardware performance~\cite{Kowert2020}, it seems important to assess whether this affects their emotional impact. And if so, what this implicates for mobile game design.
Building on promising research in other contexts as well as video games, our study seeks to further investigate the impact of color dimensions within a mobile game. Taking into consideration important influential factors such as gender, age, and experience, we aim to systematically expand on this research.
Therefore, we seek to address the following research questions:

\begin{itemize}
    \item[\textbf{RQ1}] Can the design features hue and saturation in a mobile game influence the perceived emotion of adolescents playing a mobile game significantly?
    \item[\textbf{RQ2}] Do valence and arousal ratings differ between hues that combine low and high saturation?
    \item[\textbf{RQ3}] Do gender, age, or perceived experience with video games influence valence and arousal ratings?
\end{itemize}

\section{Game Prototype}
\label{sec:prototype}

The game prototype was developed with the game engine Unity\footnote{cf. \url{https://unity.com/de}} and can be downloaded at GitHub\footnote{cf. \url{https://github.com/LeonieK/emotion-in-2d-prototype}}. 
The gameplay is reminiscent of a Mario-like jump-and-run platformer. This includes the mechanics of jumping and running as well as collecting coins both dodging fixed obstacles and moving enemies. However, it is not possible for the player to defeat the enemy. Players have five lives, and they lose one if they collide with an obstacle or enemy. Once all five lives are used up, the player is moved back to the beginning of the current level as a penalty. This gameplay was chosen because of its familiarity, as well as its reminiscence of casual game characteristics: easy to learn gameplay and simple and easy to learn controls. The player had two buttons to move left and right on the screen, and by touching the screen where there was no button, the character would jump. 

The goal of the game is to reach a Non-Playable Character (NPC) at the end of each level. The NPC asks the player for up to ten coins. Only after handing over the coins, does the player gain access to the next level.

The game consists of nine levels with increasing difficulty. The difficulty level was increased by placing obstacles and enemies in such a way that the paths became increasingly difficult to master. In addition, the jumps that the player has to perform increase in difficulty with each level.

A tileset was used for the styling of the game\footnote{cf. \url{https://assetstore.unity.com/packages/2d/environments/2d-platformer-tileset-173155}}.
We decided against adding a story to the game as this could have an impact on the perceived emotions of the players. A little motivation was provided by the introduction of the NPC asking for coins at the end of each level. This went hand-in-hand with the setting of the game. A world was created that contained man-made houses, but also plants, stones and other natural elements. A stylised comic-art style was chosen, which is relatively common especially in casual mobile games. Both the playable character as well as the NPCs have a human-like appearance.

\section{Methodology}
\label{sec:methodology}

This section outlines the methodology used in the initial study. We describe the technical background of colored game elements and specify the color stimuli used, describe the study participants, describe the questionnaires used, the study design and procedure, and the data analysis methods.

\subsection{Colored Elements}

The setup of the coloring elements was implemented with the following points in mind:
The environment and the objects of the game should remain recognisable as such. This includes that the environment is still coherent in its appearance. As an example, this means that trees retain a brown undertone for the trunk or a green undertone for the leaves, despite being colored red with high or low saturation.
In addition, objects should still remain recognisable in their functionality. This means that the coins stand out from the surroundings in any given color combination in such a way that they are recognised by the player as an object to be collected.
The character, the NPC, the enemies, as well as the coins were therefore excluded from the change of color.

\begin{figure}[ht]
    \centering
    \includegraphics[width=0.48\textwidth]{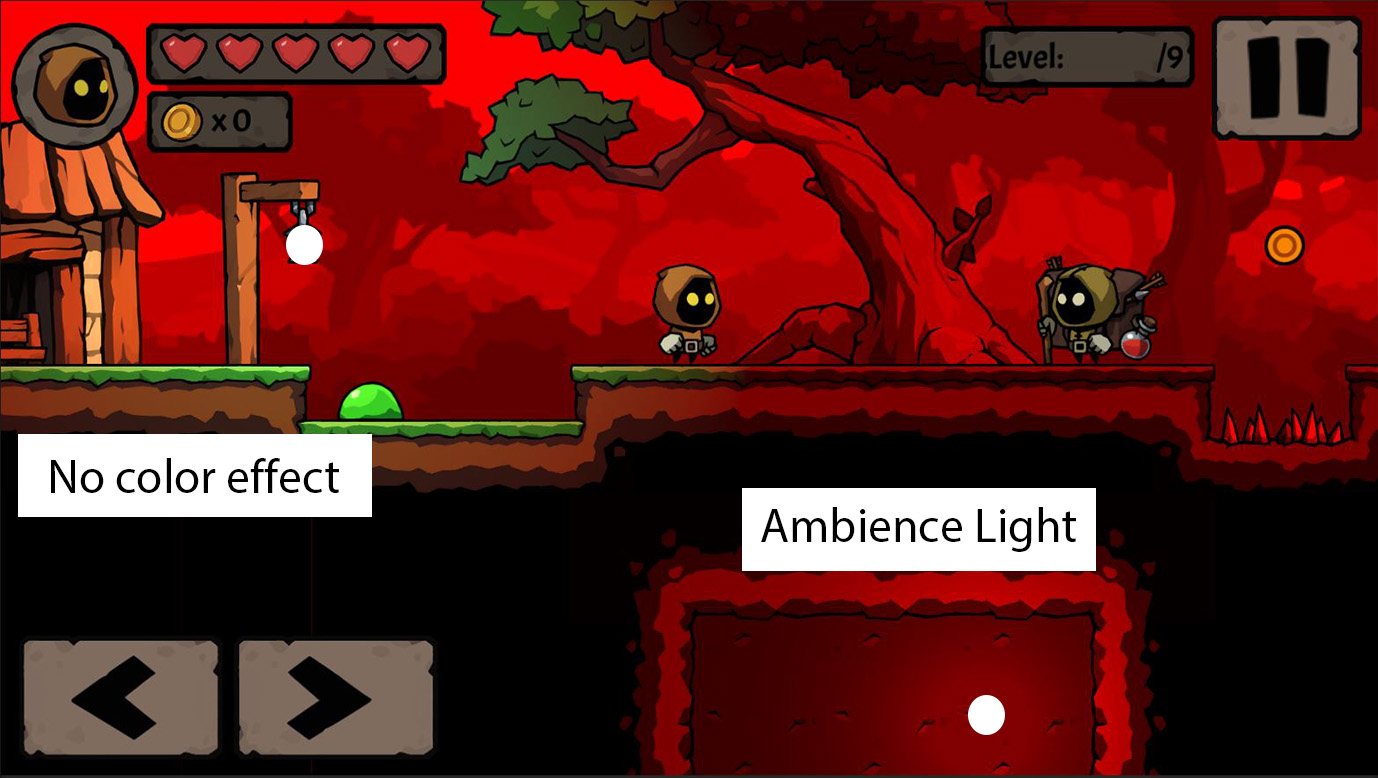}
    \caption{Screenshot of (a part of) one game level in hue red. Light sources are marked with a white dot. The light source labeled "No Color Effect" applies a white light to the environment to balance some intense color effects. "Ambience Lights" apply color-specific lighting, in this case red.}
    \Description{Screenshot of (a part of) one game level in hue red. Light sources are marked with a white dot. The light source labeled "No Color Effect" applies a white light to the environment to balance some intense color effects. "Ambience Lights" apply color-specific lighting, in this case red.}
    \label{fig:level}
\end{figure}

Technically, this was implemented using Lights 2d\footnote{cf. \url{https://docs.unity3d.com/Packages/com.unity.render-pipelines.universal@7.1/manual/Lights-2D-intro.html}} provided by Unity.
A global light with an intensity of 1.5 was applied to the background layer, as well as to all decorative elements and the environment, with a blend style of 0. Only the hue and saturation were adjusted, not the brightness.
Point lights were used to add ambience and enhance the effect of the color change. With the intensity set to 2 and the blend style set to multiply, the brightness in these areas was increased to some extent. Light sources without color effects were also used to keep the levels from looking too unnatural (cf. Fig.~\ref{fig:level}).

\subsection{Participants}
\label{sec:metho_participants}

Twenty-three volunteers were recruited. Prior to the study, each participant was required to complete the Ishihara~\cite{Ishihara1987} color blindness test to determine normal color vision. Eleven plates were displayed on the screen of their mobile phone - if they could read at least seven, they were considered to have normal color vision. 
Data from two participants had to be excluded because they failed the Ishihara color blindness test. These data have been excluded from the following statements.
The mean age of participants was 16.7 years ranging from 14 to 20 years in age. Among the participants, 8 (38.1\%) persons identified themselves as female while 13 (61.9\%) persons did so as male.

Participation in the study was voluntary. Each participant had to confirm that they understood the general conditions before starting the study. All data collected during the study was non-personal.
The participants were not informed about the research questions in detail. However, they were told at the beginning that they would be asked about their emotional state while playing the game.

Participants used their own mobile phones to ensure familiarity with their operation.

\subsection{Questionnaires}

Participants were presented with different types of questionnaires. A general questionnaire asked for age, gender, and experience in games. 
The experience in games could be estimated based on three categories. As the participants were German, the categories were described to them in German. 
The categories are listed below in their original German language with an English translation. 
\begin{itemize}
\item Low Experience: “Ich kenne mich nicht aus oder spiele kaum bis gar nicht” (I don’t know anything about games or I hardly play at all). 
\item Medium Experience: “Ich kenne mich etwas aus oder spiele manchmal” (I know something about games or play sometimes). 
\item High Experience: “Ich kenne mich aus oder spiele regelmäßig” (I know about games and play regularly)
\end{itemize}
To check for any color deficiencies, the Ishihara color blind test was used. (see Sec.~\ref{sec:metho_participants} above)

\begin{figure}[ht]
    \centering
    \includegraphics[width=0.48\textwidth]{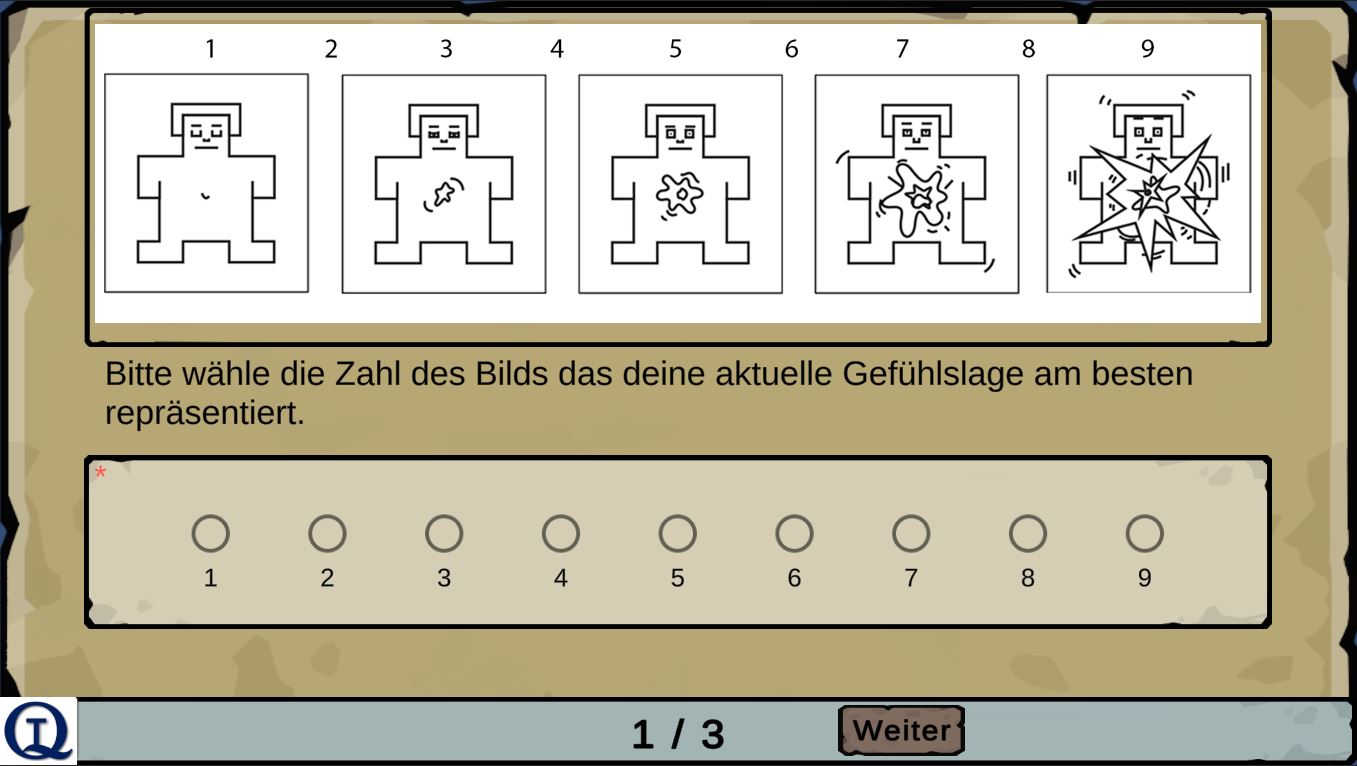}
    \caption{Screenshot of the in-game view of the SAM questionnaire using arousal as an example. Participants were asked in German (English translation by the authors): “Please choose the number of the picture that best represents your current emotional state".}
    \Description{Screenshot of the in-game view of the SAM questionnaire using arousal as an example. Participants were asked in German (English translation by the authors): “Please choose the number of the picture that best represents your current emotional state".}
    \label{fig:sam}
\end{figure}

The Self-Assessment Manikin (SAM) questionnaire~\cite{Bradley1994} was used to assess participants' current emotional state. The SAM provides a simple, nonverbal, pictorial way to assess the affective dimensions of valence, arousal, and dominance. 
Participants are shown five pictures each, from which they select one that best represents their emotional state. The score ranges from 1 to 9, which means that it is possible to settle between two pictures. Figure~\ref{fig:sam} shows the version of the questionnaire shown to the participants in the game.

Arousal and valence are the primary dimensions of the SAM questionnaire used to identify emotions. Dominance is used to index whether participants feel in control of the game. A score of 9 on Dominance indicates a feeling of maximum control.

All questionnaires were integrated into the game prototype. This was done to ensure that the participants remained somewhat immersed while answering the questions. The Asset Questionnaire Toolkit~\footnote{\url{https://assetstore.unity.com/packages/tools/gui/questionnairetoolkit-157330}} was used to create all questionnaires. (cf.~\cite{Jansen2019})

\subsection{Color Stimuli}

Nine different colors were chosen for integration into the game. Four hues (red, blue, green, and yellow) were combined with high and low saturation, with the brightness set to high. A neutral gray color with high brightness was included as well.
The selection of the four different hues was based on their reported effects on arousal and valence. Red as well as high saturation is reported to reach strong effects on arousal as well as valence~\cite{Joosten2012, Elliot2019}. Green and blue seem to achieve a transition from high to low arousal, whereby blue achieved high effects for valence~\cite{Wilms2018} and lead to calmness or relaxation~\cite{Elliot2019}. Yellow is reported to reach a strong effect on valence~\cite{Joosten2012}.

Our study initially implemented a 4x2x2 factorial design that manipulated saturation and brightness at both low and high levels. Due to the anticipated number of participants, we decided to eliminate brightness variation to simplify the study design. This is because color research indicates that saturation has a greater impact on affect~\cite{Elliot2019}.
The final selection of hue and saturation with RGB values is shown in table~\ref{tab:colors}.

\begin{table}[h]
\caption{Combinations of hue and saturation with RGB values as used in the investigation.}
\Description{Combinations of hue and saturation with RGB values as used in the investigation.}
\label{tab:colors}
\begin{tabular}{l|c|l}
\toprule
hue & saturation & hue + saturation in RGB \\
\midrule
blue & high & 0, 110, 255 \\
& low & 111, 116, 157 \\
green & high & 0, 149, 0 \\
& low & 87, 128, 97 \\
red & high & 246, 0, 0 \\
& low & 158, 106, 93 \\
yellow & high & 255, 255, 0 \\
& low & 234, 228, 176 \\
neutral & low & 121, 118, 119 \\
\bottomrule
\end{tabular}
\end{table}
\subsection{Design and Procedure}

The study employed a within-subjects design with hue and saturation as independent variables and SAM rating as the dependent variable. Each participant played a game consisting of 9 main levels, each featuring a unique color combination randomly assigned to the level.

Before starting the game, participants received a brief introduction that explained the topic of the study and the data-handling procedure. Additionally, we verified that the game was correctly configured on the mobile browser to ensure full-screen gameplay.

At the start of the game, participants gave informed consent to be part of the study on a voluntary basis. They were also informed that they could withdraw from the study at any time without facing any negative consequences.

\begin{figure*}[ht]
    \centering
    \includegraphics[width=0.8\textwidth]{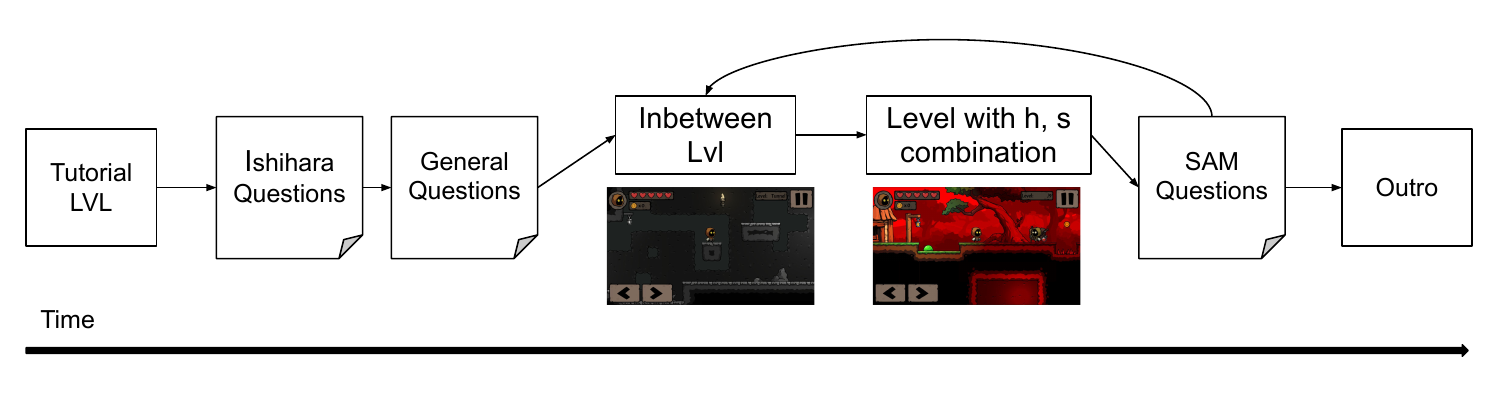}
    \caption{Structural illustration of the process of the study.}
    \Description{Structural illustration of the process of the study.}
    \label{fig:procedure}
\end{figure*}

The process after the start of the game is shown in Figure~\ref{fig:procedure}. After pressing “Start Game”, participants reach the first level functioning as a tutorial. Here, participants familiarize themselves with the game's mechanics. The coloring is neutral with low saturation.
After successfully completing the introductory level, the player will speak with an NPC who will ask questions about the Ishihara test plates, age, gender, and gaming experience.

The participant is placed by the NPC in an neutral in-between level, before the main level starts. After each main level, completion of the SAM questionnaire is required. Participants then progress through another in-between level. 
Once the survey is finished participants are invited to provide anonymous feedback.

\subsection{Data Analysis}

A two-way repeated measures analysis of variance (rm ANOVA) was conducted to examine the effects of hue and saturation as within-factors. The dataset was divided according to age groups and gender, and analyzed using a two-way rm ANOVA. 
To incorporate the neutral color value, a one-way rm ANOVA was performed with hue as the within-subject factor.

Data analysis was performed using python with the NumPy, pandas, matplotlib and pingouin packages. Pingouin v0.5.2~\cite{Vallat2018} was used to perform all rm ANOVAs.

\section{Results}
\label{sec:results}
In the following, we present the results of our data analysis, discussing first general findings and then detailed findings for the measures of gender, age, video game experience, and time spent in each game level.

\begin{figure*}[ht]
    \centering
    \includegraphics[width=0.6\textwidth]{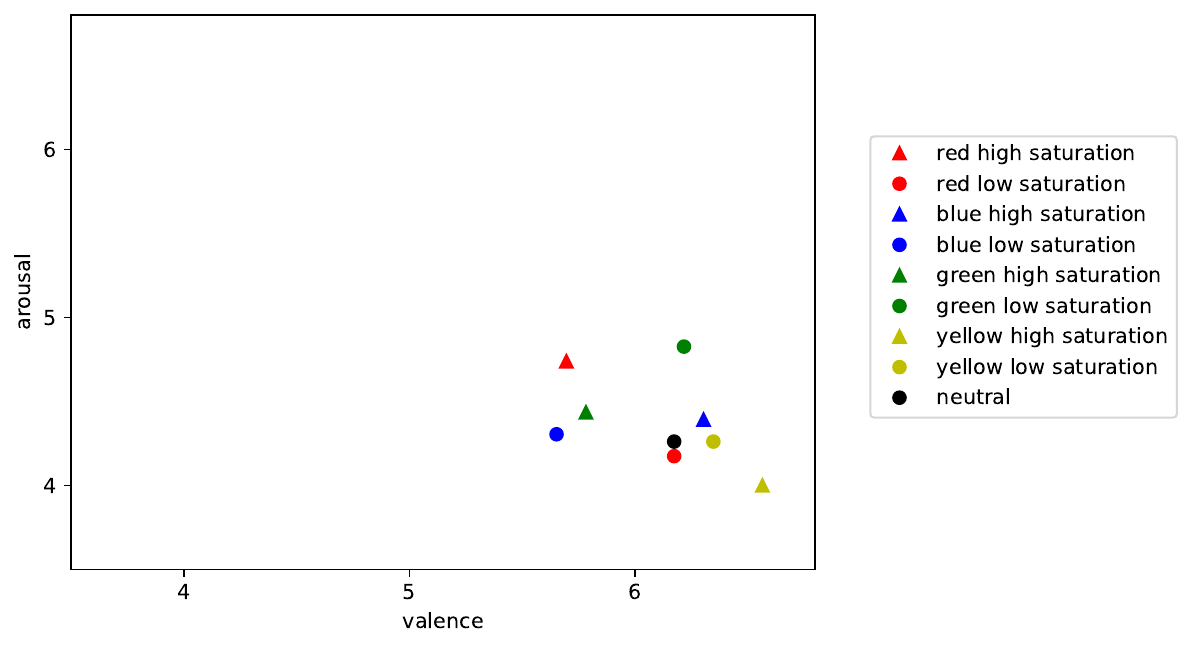}
    \caption{Mean values of each hue and saturation combination for valence and arousal. The hues of a point indicate the corresponding hue, a triangle indicates high, a dot indicates low saturation. For better readability of the plots, we have truncated the axis at 3.5 and 6.8, respectively.}
    \Description{Mean values of each hue and saturation combination for valence and arousal. The hues of a point indicate the corresponding hue, a triangle indicates high, a dot indicates low saturation. For better readability of the plots, we have truncated the axis at 3.5 and 6.8, respectively.}
    \label{fig:valenceXarousal}
\end{figure*}

The SAM ratings for the whole dataset indicated little variation among the different color stimuli. A cluster in the coordinate system's lower-right range emerges from the average values for arousal and valence (cf. Figure~\ref{fig:valenceXarousal}). This shows that the ratings are generally close to the neutral rating of $5$. 
This is a low variation compared to the results of Wilms and Oberfeld~\cite{Wilms2018}, but similar to the results reported by Joosten et al.~\cite{Joosten2012}.

A two-way rm ANOVA~\cite{Girden1992} was performed for the arousal, valence, and dominance ratings using the within-factors hue and saturation. No significant effect could be determined here. To include the gray level, a one-way rm ANOVA was conducted, considering the within-factor variable of hue. However, no significant effect was found in this case either.
Thus, it can be concluded that neither hue nor saturation significantly affected the ratings of arousal, valence, and dominance.

\subsection{Gender}
\label{sec:gender}

Given the relevance of gender in video games and color perception, we hypothesized that participants identifying as different gender would show different responses. Amongst the participants there were no persons identifying as non-binary. As such, we conducted a one-way rm ANOVA with the within-factor hue to explore this notion for male and female participatns. However, no significant effect was observed.

\begin{figure*}[ht]
    \centering
    \includegraphics[width=0.45\textwidth]{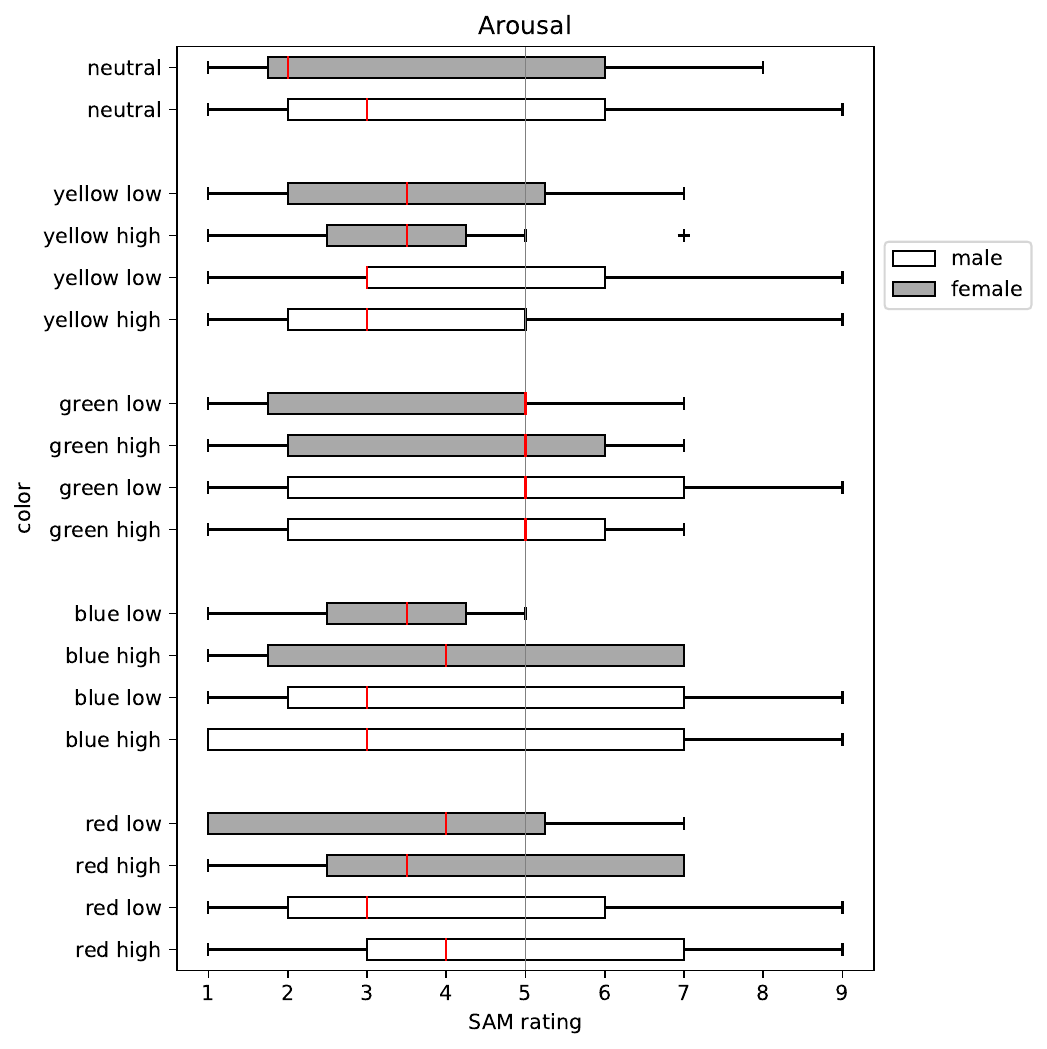}
    \includegraphics[width=0.45\textwidth]{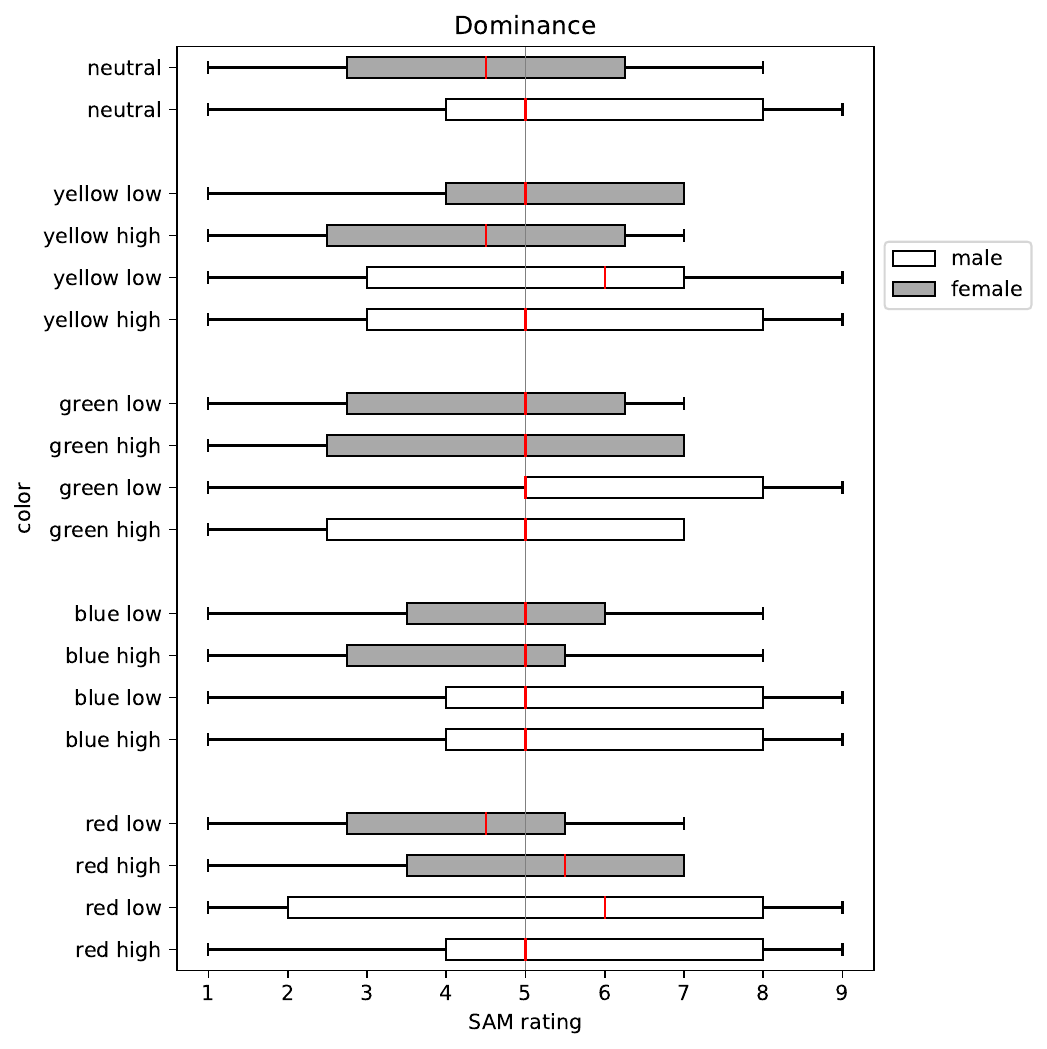}
    \includegraphics[width=0.45\textwidth]{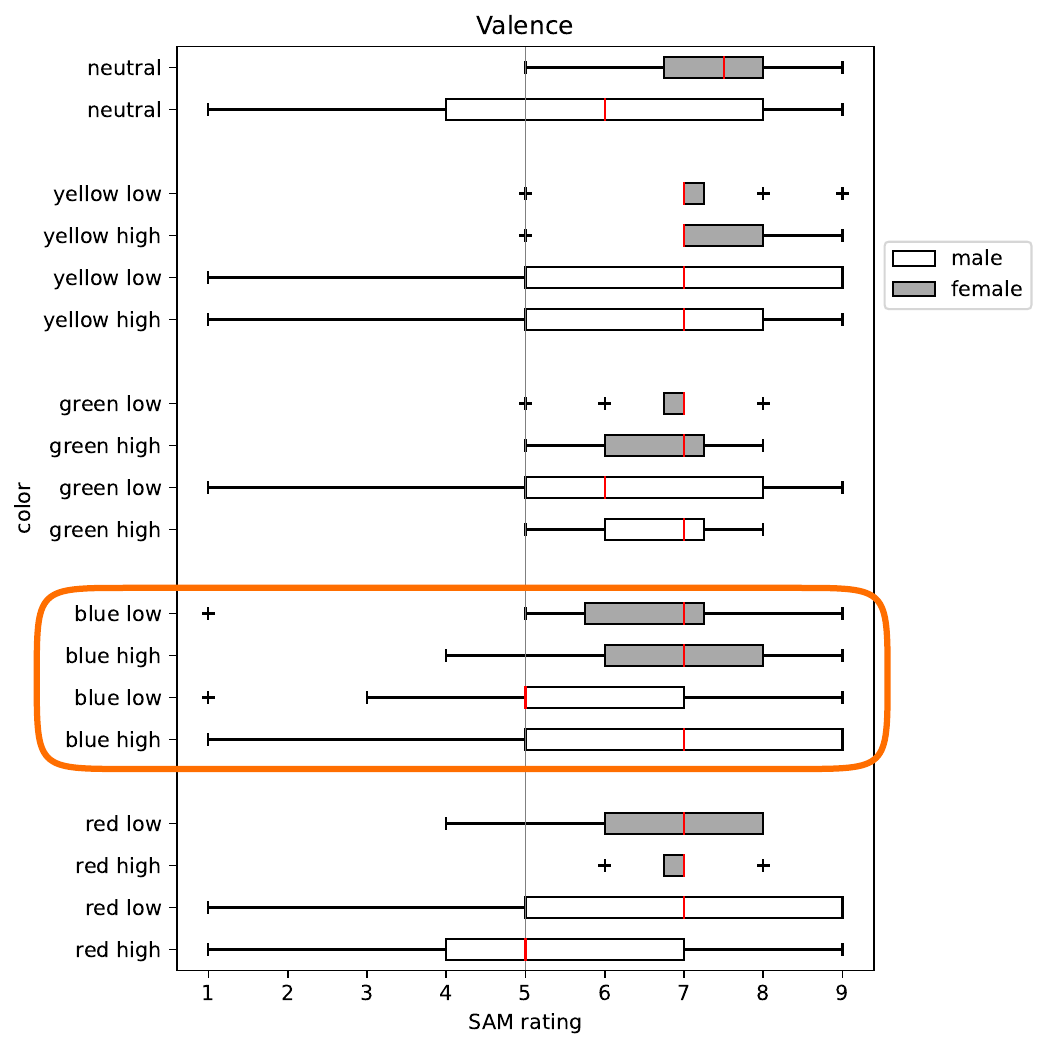}
    \caption{Boxplots of the results received for arousal (upper left), dominance (upper right)), and valence (lower) from all participants. Boxplots are sorted according to the hue, saturation, and gender. Within the valence figure, the values with a significant difference were highlighted with an orange box. The median is indicated in red, outliers are marked with a ‘+’ symbol.}
    \Description{Boxplots of the results received for arousal (upper left), dominance (upper right)), and valence (lower) from all participants. Boxplots are sorted according to hue, saturation, and gender of the corresponding participant. Within the valence figure, the values with a significant difference were highlighted with an orange box.}
    \label{fig:boxplots}
\end{figure*}

In addition, we conducted a two-way rm ANOVA using within-factors hue and saturation, separately for male and female participants.
A significant effect on valence ratings was found on the interaction between hue and saturation for male participants ($F(3,36) = 3.129$, $p = 0.038 (<.05)$, $np2 = 0.207$).
Post-hoc tests (pairwise t-tests) showed a significant effect for the hue blue with low and high saturation ($T(12) = -2.144$, $p = 0.053 (two-tailed)$). Figure~\ref{fig:boxplots} shows boxplots for the SAM ratings for arousal, valence, and dominance. The significant effect for valence ratings is indicated by the difference in the median marked in the lower part of Figure~\ref{fig:boxplots}. 
For arousal the interquartile ranges are mostly below 7. Neutral color within females has the lowest median with a rating of 2. Green with low and high saturation for male and female has the highest with a rating of 5. This also indicates the lowest difference between male and female responses for green.
Regarding valence the interquartile ranges are almost all above the SAM rating 5. Blue low within male and red high within male participants is the lowest median with a rating of 5.
For dominance the interquartile ranges are mostly clustered in the middle of SAM ratings. The lowest median is 4.5 for neutral and red low within female participants. The highest median is the SAM rating 6 for red low and yellow low within male participants.

\begin{figure*}[ht]
    \centering
    \includegraphics[width=0.5\textwidth]{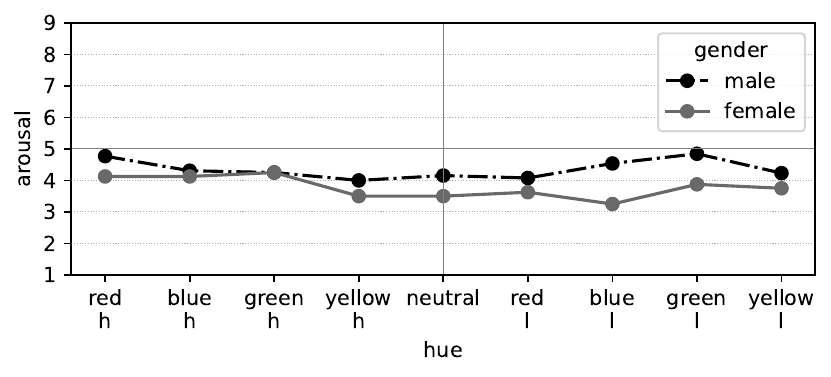}
    \includegraphics[width=0.5\textwidth]{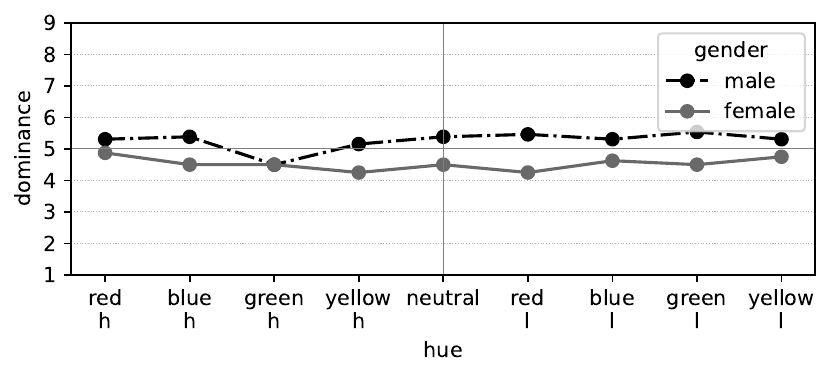}
    \includegraphics[width=0.5\textwidth]{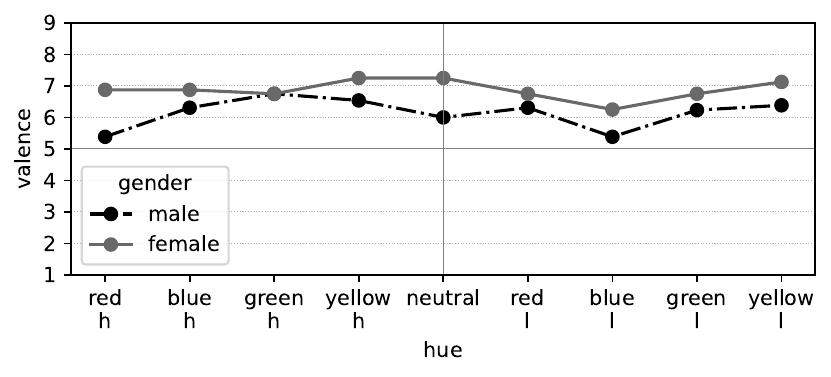}
    \caption{Plots of arousal (upper), dominance (middle), and valence (lower) mean values against hues with lower as well as higher saturation. Depicted are the mean values received for female and male participants. High saturation results are shown to the left, low saturation results to the right of the neutral gray in the middle of each figure.}
    \Description{Plots of arousal (upper), dominance (middle), and valence (lower) mean values against hues with lower as well as higher saturation. Depicted are the mean values received for female and male participants. High saturation results are shown to the left, low saturation results to the right of the neutral gray in the middle of each figure.}
    \label{fig:huesXplot}
\end{figure*}

Figure~\ref{fig:huesXplot} clearly shows that there is an overall difference between the SAM ratings of male and female participants. The difference is stronger for high saturation. Male participants reported higher arousal on average, although this never exceeded 5. Male participants also generally reported higher ratings for dominance, with values here ranging around 5. For valence, female participants generally gave higher ratings, with all averages remaining above 5. The averages for green with high saturation are the same for male and female participants.

In summary, there was a difference between male and female participants in that arousal and dominance were perceived higher by males and valence was perceived higher by females. In addition, a significant effect was found for male participants for the interaction of blue hue with low and high saturation.

\subsection{Experience}
\label{sec:experience}

We hypothesized that there is a difference between participants who rated their experience with video games as low, medium, or high. 
A one-way rm ANOVA with the within-factor hue was conducted; again no significant effect was found.
Furthermore, a two-way rm ANOVA was conducted for either of the three experience levels with the within-factors hue and saturation.
Among participants that estimated their game experience as high, a significant effect of the interaction between hue and saturation on valence ratings can be reported ($F(3,18) = 4$, $p = 0.024 (<.05)$, $np2 = 0.4$). Post-hoc tests (pairwise t-tests) showed no further significant results.

All of the participants who rated their experience as high are male $(n=7)$. Thus, we consider this to be a consequence of the results shown for male participants and the corresponding valence ratings.

\subsection{Age}
\label{sec:age}

We hypothesized that there is a difference in ratings between groups of participants of different age. For the investigation, we divided the participants at the mean $16.7$, resulting in two groups: $14$ to $16$ year olds and $17$ to $20$ year olds.
A two-way rm ANOVA was conducted for arousal, valence, and dominance with the within-factors hue and saturation. No significant effect in either of the groups of participants was found.

Additionally, a one-way rm ANOVA  was performed with hue as the within factor. Again, no significant effect was found.

\subsection{Time}
\label{sec:time}

\begin{figure}[ht]
    \centering
    \includegraphics[width=0.35\textwidth]{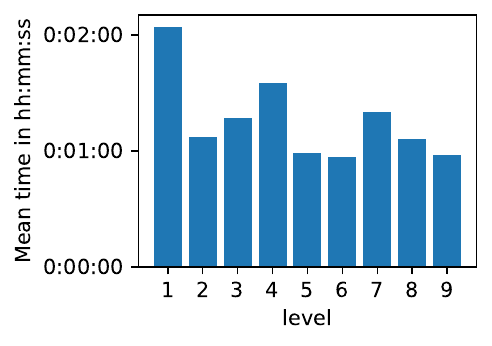}
    \caption{Bar chart showing the mean time in hh:mm:ss for each level.}
    \Description{Bar chart showing the mean time in hh:mm:ss for each level.}
    \label{fig:time}
\end{figure}

We also considered how much time participants spent, on average, in each level. Figure~\ref{fig:time} shows that the time spent in a level generally decreased towards the higher levels. It is interesting that the time spent in the first level is considerably greater compared to the other levels.

\section{Analysis and Discussion}
\label{sec:analysis}

In this section, we discuss the results of the study conducted. In doing so we also address research questions RQ1, RQ2, and RQ3. (cf. Sec.~\ref{sec:intro})

With respect to RQ1(do hue and saturation influence perceived emotion) and RQ2(do valence and arousal ratings differ between combinations of hue and saturation), we found that neither hue nor saturation significantly affected arousal, valence, and dominance ratings. There are several possible explanations for this lack of significant effect.
On the one hand, these results may indicate that the integration of color through the use of tinted lighting in the game character's environment does not have the desired effect on players. A lot of emphasis was placed on maintaining the overall sense of a somewhat realistic game environment. Considering that Plass et al. found an effect when changing the color of a character\cite{Plass2020}, it may have helped to change the color of the characters as well. 
Game elements other than color did not change significantly during gameplay. Only a slight increase in difficulty was added to prevent players from getting bored. Another indication that the color changes did not influence players to have different emotions is the cluster of mean scores depicted in figure~\ref{fig:valenceXarousal}. 
Given these results, it is possible that color alone is not an effective measure for eliciting emotion in mobile games.

On the other hand, these results may indicate that the use of color on mobile phones is not effective given the screen sizes, where the color changes may have been lost. This particularly holds for the research done on larger screens~\cite{Wilms2018, Joosten2012, Geslin2016, Elliot2019} or other media~\cite{Elliot2019} that found significant effects of color on emotion. Research comparing smartphone screen sizes also suggests that perceived affective quality is higher when using large screens~\cite{Kim2014}.

Players spent an average of 1:16 minutes in each color condition. Given the recognition that time has an important impact on emotional responses~\cite{Landers1980}, this may have been a too short period of time to elicit an emotional response. This makes time a factor that deserves further investigation, as many mobile games are designed to be short lived~\cite{Anable2018}.
As described in section "Time"~(\ref{sec:time}), most time was spent in the first level. The average short time spent in general and the noticeable drop in time spent in higher levels could indicate that the participants were not able to immerse into the game. This could have affected the intensity of the emotions perceived. Jennett et al.~\cite{Jennett2008} state that immersion, or the lack of it can be evident by the time spent. Another explanation, at least for the long time spent in the first level, is that players needed to get used to the game. We put a tutorial level before this first level, which may have been too short to allow players to become familiar enough with the game as a whole.

Related to RQ3 (do gender, age, or perceived experience influence valence and arousal ratings, cf. Sec.~\ref{sec:intro}), we found a significant effect within male participants for the interaction between hue and saturation in valence ratings. In particular the hue blue with low and high saturation had a significant effect. At this point we find our results in line with the findings of Wilms and Oberfeld~\cite{Wilms2018} who also reported an interaction of hue and gender as well as saturation and gender. 
Considering that color preference research has shown that blue is preferred by males~\cite{Elliot2019}, this finding may suggest that attractiveness has an effect on valence or the pleasantness experienced.

We also observed a general difference between the submitted SAM ratings of male and female participants.   
Male participants generally experienced higher arousal, which could be explained by the genre of the game and the corresponding gender differences. The genre is a platformer, that tends to be more challenging, which is preferred by men, as opposed to women's preference for immersive and relaxing games~\cite{Tondello2019}.

To fully include RQ3, we briefly review our observations related to the differences in experience with games as well.
We found a significant effect for male participants with highest experience on the interaction between hue and saturation on valence. We explain this as a consequence of the significant effect that arose for valence within all male participants. However, it is interesting that there is a significant effect for participants with a high experience. This differs from Joosten et al.~\cite{Joosten2012}, who found strong effects only for individuals with little experience.

In the following, we also discuss limitations of our study. We hope this will be helpful for contextualizing our findings as well as future studies in this area.
Given the number of participants, it would have been helpful to obtain ethnographic observations in addition to questionnaire responses. We provided an optional feedback box for participants to fill out, but we did not receive any useful responses on this. It would have been helpful to use a method like "Think Aloud"~\cite{Charters2003} to get participants talking about their current emotional state and the game.

Culture is an important element in how colors are perceived and what color preferences exist. We did not ask the participants about their cultural background as they were all German speaking. In hindsight, however, it would have been interesting to see if differences in culture affected the results.

Since mobile games are played in different environments, we tried to make the survey as realistic as possible by conducting it mostly outdoors and letting participants use their own phones. This may have affected the visibility of the game on the screen, as well as the color, since different phones may have different display settings.

The game had a neutral level (gray colors) at the beginning of the study, with the intention that players would reach a relatively neutral emotional state.  However, the initial emotional state was not assessed, and we cannot compare what the emotional state was at the beginning of the first level with the studied color stimuli.

\section{Conclusion and future work}
\label{sec:conclusion}
In this paper we investigated how color stimuli can influence emotions in mobile games. 
Understanding how we can create emotion through design features can be an important tool for game designers, as well as for educational game designers to facilitate the design process. Emotional design is a line of research that addresses this issue on a broader scale. The use of color is promising, given the large body of evidence for its effectiveness in other contexts. The context of mobile games is also important as these devices have different requirements than other video games.

We reviewed literature related to color, video games, mobile games, and emotions. From this, we hypothesized that the design features of hue and saturation in a mobile game influence adolescents' perceived emotion; that there are interactions in these features; and that gender, age, and perceived video game experience influence them. 

We conducted an initial study to address this hypothesis with the following results:
We found that neither hue nor saturation significantly affected arousal, valence, and dominance ratings, which may indicate that color alone is not an effective measure for eliciting emotion in mobile games. Explanations for this could be that mobile games have small screen sizes, imply short-lived gameplay, and provide a context that almost presupposes distractions. 
Further research into time spent on mobile games and the relevance of screen size to emotional design would help understand if the use of color needs to be improved or if we need additional emotional design elements to evoke emotion.

We have observed an effect of gender on emotional responses. Here the interaction of hue and saturation among male participants for valence ratings was remarkable as well as the interaction of hue and saturation among male participants with high video game experience. This may be an indication that color preference influences perceived pleasantness. 
Building on these findings, it would be helpful to see if these effects can be reproduced for women. Another interesting follow-up study would be to ask about color preferences and compare perceived emotions while playing in environments with preferred and non-preferred colors.

\bibliographystyle{ACM-Reference-Format}
\bibliography{sample-base}


\begin{thebibliography}{36}


\ifx \showCODEN    \undefined \def \showCODEN     #1{\unskip}     \fi
\ifx \showDOI      \undefined \def \showDOI       #1{#1}\fi
\ifx \showISBNx    \undefined \def \showISBNx     #1{\unskip}     \fi
\ifx \showISBNxiii \undefined \def \showISBNxiii  #1{\unskip}     \fi
\ifx \showISSN     \undefined \def \showISSN      #1{\unskip}     \fi
\ifx \showLCCN     \undefined \def \showLCCN      #1{\unskip}     \fi
\ifx \shownote     \undefined \def \shownote      #1{#1}          \fi
\ifx \showarticletitle \undefined \def \showarticletitle #1{#1}   \fi
\ifx \showURL      \undefined \def \showURL       {\relax}        \fi
\providecommand\bibfield[2]{#2}
\providecommand\bibinfo[2]{#2}
\providecommand\natexlab[1]{#1}
\providecommand\showeprint[2][]{arXiv:#2}

\bibitem[Anable(2018)]%
        {Anable2018}
\bibfield{author}{\bibinfo{person}{Aubrey Anable}.}
  \bibinfo{year}{2018}\natexlab{}.
\newblock \bibinfo{booktitle}{\emph{Playing With Feelings: Video Games and
  Affect}}.
\newblock \bibinfo{publisher}{University of Minnesota Press}.
\newblock


\bibitem[Aslam(2006)]%
        {Aslam2006}
\bibfield{author}{\bibinfo{person}{Mubeen~M. Aslam}.}
  \bibinfo{year}{2006}\natexlab{}.
\newblock \showarticletitle{Are you selling the right colour? A cross-cultural
  review of colour as a marketing cue}.
\newblock \bibinfo{journal}{\emph{Journal of Marketing Communications}}
  \bibinfo{volume}{12} (\bibinfo{date}{3} \bibinfo{year}{2006}),
  \bibinfo{pages}{15--30}.
\newblock
Issue 1.
\showISSN{13527266}
\urldef\tempurl%
\url{https://doi.org/10.1080/13527260500247827}
\showDOI{\tempurl}


\bibitem[Bradley and Lang(1994)]%
        {Bradley1994}
\bibfield{author}{\bibinfo{person}{Margaret~M Bradley} {and}
  \bibinfo{person}{Peter~J Lang}.} \bibinfo{year}{1994}\natexlab{}.
\newblock \showarticletitle{Measuring emotion: the self-assessment manikin and
  the semantic differential}.
\newblock \bibinfo{journal}{\emph{Journal of behavior therapy and experimental
  psychiatry}} \bibinfo{volume}{25}, \bibinfo{number}{1}
  (\bibinfo{year}{1994}), \bibinfo{pages}{49--59}.
\newblock


\bibitem[Brom et~al\mbox{.}(2018)]%
        {Brom2018}
\bibfield{author}{\bibinfo{person}{Cyril Brom}, \bibinfo{person}{Tereza
  Stárková}, {and} \bibinfo{person}{Sidney~K. D'Mello}.}
  \bibinfo{year}{2018}\natexlab{}.
\newblock \bibinfo{title}{How effective is emotional design? A meta-analysis on
  facial anthropomorphisms and pleasant colors during multimedia learning}.
\newblock , \bibinfo{numpages}{100-119}~pages.
\newblock
\showISSN{1747938X}
\urldef\tempurl%
\url{https://doi.org/10.1016/j.edurev.2018.09.004}
\showDOI{\tempurl}


\bibitem[Chang and Hwang(2019)]%
        {Chang2019}
\bibfield{author}{\bibinfo{person}{Ching~Yi Chang} {and}
  \bibinfo{person}{Gwo~Jen Hwang}.} \bibinfo{year}{2019}\natexlab{}.
\newblock \showarticletitle{Trends in digital game-based learning in the mobile
  era: A systematic review of journal publications from 2007 to 2016}.
\newblock \bibinfo{journal}{\emph{International Journal of Mobile Learning and
  Organisation}}  \bibinfo{volume}{13} (\bibinfo{year}{2019}),
  \bibinfo{pages}{68--90}.
\newblock
Issue 1.
\showISSN{17467268}
\urldef\tempurl%
\url{https://doi.org/10.1504/IJMLO.2019.096468}
\showDOI{\tempurl}


\bibitem[Charters(2003)]%
        {Charters2003}
\bibfield{author}{\bibinfo{person}{Elizabeth Charters}.}
  \bibinfo{year}{2003}\natexlab{}.
\newblock \showarticletitle{The Use of Think-aloud Methods in Qualitative
  Research An Introduction to Think-aloud Methods}.
\newblock \bibinfo{journal}{\emph{Brock Education Journal}}
  \bibinfo{volume}{12} (\bibinfo{date}{7} \bibinfo{year}{2003}).
\newblock
Issue 2.
\urldef\tempurl%
\url{https://doi.org/10.26522/brocked.v12i2.38}
\showDOI{\tempurl}


\bibitem[Ekman(2004)]%
        {Ekman2004}
\bibfield{author}{\bibinfo{person}{Paul Ekman}.}
  \bibinfo{year}{2004}\natexlab{}.
\newblock \showarticletitle{Emotions revealed}.
\newblock \bibinfo{journal}{\emph{Bmj}} \bibinfo{volume}{328},
  \bibinfo{number}{Suppl S5} (\bibinfo{year}{2004}).
\newblock


\bibitem[Elliot(2019)]%
        {Elliot2019}
\bibfield{author}{\bibinfo{person}{Andrew~J. Elliot}.}
  \bibinfo{year}{2019}\natexlab{}.
\newblock \showarticletitle{A Historically Based Review of Empirical Work on
  Color and Psychological Functioning: Content, Methods, and Recommendations
  for Future Research}.
\newblock \bibinfo{journal}{\emph{Review of General Psychology}}
  \bibinfo{volume}{23} (\bibinfo{date}{6} \bibinfo{year}{2019}),
  \bibinfo{pages}{177--200}.
\newblock
Issue 2.
\showISSN{19391552}
\urldef\tempurl%
\url{https://doi.org/10.1037/gpr0000170}
\showDOI{\tempurl}


\bibitem[Elliot and Maier(2012)]%
        {Elliot2012}
\bibfield{author}{\bibinfo{person}{Andrew~J. Elliot} {and}
  \bibinfo{person}{Markus~A. Maier}.} \bibinfo{year}{2012}\natexlab{}.
\newblock \showarticletitle{Chapter two - Color-in-Context Theory}.
\newblock \bibinfo{series}{Advances in Experimental Social Psychology},
  Vol.~\bibinfo{volume}{45}. \bibinfo{publisher}{Academic Press},
  \bibinfo{pages}{61--125}.
\newblock
\showISSN{0065-2601}
\urldef\tempurl%
\url{https://doi.org/10.1016/B978-0-12-394286-9.00002-0}
\showDOI{\tempurl}


\bibitem[Fairchild(2015)]%
        {Fairchild2015}
\bibfield{author}{\bibinfo{person}{Mark~D. Fairchild}.}
  \bibinfo{year}{2015}\natexlab{}.
\newblock \showarticletitle{Color models and systems}.
\newblock \bibinfo{publisher}{Cambridge University Press},
  \bibinfo{pages}{9--26}.
\newblock
\urldef\tempurl%
\url{https://doi.org/10.1017/cbo9781107337930.003}
\showDOI{\tempurl}


\bibitem[Frome(2007)]%
        {Frome2007}
\bibfield{author}{\bibinfo{person}{Jonathan Frome}.}
  \bibinfo{year}{2007}\natexlab{}.
\newblock \showarticletitle{Eight Ways Videogames Generate Emotion.}. In
  \bibinfo{booktitle}{\emph{DiGRA conference}}. \bibinfo{pages}{831--835}.
\newblock


\bibitem[Geslin et~al\mbox{.}(2016)]%
        {Geslin2016}
\bibfield{author}{\bibinfo{person}{Erik Geslin}, \bibinfo{person}{Laurent
  Jégou}, {and} \bibinfo{person}{Danny Beaudoin}.}
  \bibinfo{year}{2016}\natexlab{}.
\newblock \showarticletitle{How Color Properties Can Be Used to Elicit Emotions
  in Video Games}.
\newblock \bibinfo{journal}{\emph{International Journal of Computer Games
  Technology}} (\bibinfo{year}{2016}).
\newblock
\showISSN{16877055}
\urldef\tempurl%
\url{https://doi.org/10.1155/2016/5182768}
\showDOI{\tempurl}


\bibitem[Girden(1992)]%
        {Girden1992}
\bibfield{author}{\bibinfo{person}{Ellen~R. Girden}.}
  \bibinfo{year}{1992}\natexlab{}.
\newblock \bibinfo{booktitle}{\emph{ANOVA: Repeated measures.}}
\newblock \bibinfo{publisher}{Sage Publications, Inc},
  \bibinfo{address}{Thousand Oaks, CA, US}. vi, 77--vi, 77 pages.
\newblock
\showISBNx{0-8039-4257-5 (Paperback)}


\bibitem[Greenberg et~al\mbox{.}(2010)]%
        {Greenberg2010}
\bibfield{author}{\bibinfo{person}{Bradley~S. Greenberg}, \bibinfo{person}{John
  Sherry}, \bibinfo{person}{Kenneth Lachlan}, \bibinfo{person}{Kristen Lucas},
  {and} \bibinfo{person}{Amanda Holmstrom}.} \bibinfo{year}{2010}\natexlab{}.
\newblock \showarticletitle{Orientations to video games among gender and age
  groups}.
\newblock \bibinfo{journal}{\emph{Simulation and Gaming}}  \bibinfo{volume}{41}
  (\bibinfo{date}{4} \bibinfo{year}{2010}), \bibinfo{pages}{238--259}.
\newblock
Issue 2.
\showISSN{10468781}
\urldef\tempurl%
\url{https://doi.org/10.1177/1046878108319930}
\showDOI{\tempurl}


\bibitem[Günther and Stiles(2000)]%
        {Wyszecki2000}
\bibfield{author}{\bibinfo{person}{Wyszecki Günther} {and}
  \bibinfo{person}{W.~S. Stiles}.} \bibinfo{year}{2000}\natexlab{}.
\newblock \bibinfo{booktitle}{\emph{Color Science: Concepts and Methods,
  Quantitative Data and Formulae}}.
\newblock \bibinfo{publisher}{Wiley-Interscience}.
\newblock
\showISBNx{978-0471399186}


\bibitem[Hemenover and Bowman(2018)]%
        {Hemenover2018}
\bibfield{author}{\bibinfo{person}{Scott~H. Hemenover} {and}
  \bibinfo{person}{Nicholas~D. Bowman}.} \bibinfo{year}{2018}\natexlab{}.
\newblock \showarticletitle{Video games, emotion, and emotion regulation:
  expanding the scope}.
\newblock \bibinfo{journal}{\emph{Annals of the International Communication
  Association}}  \bibinfo{volume}{42} (\bibinfo{date}{4} \bibinfo{year}{2018}),
  \bibinfo{pages}{125--143}.
\newblock
Issue 2.
\showISSN{2380-8985}
\urldef\tempurl%
\url{https://doi.org/10.1080/23808985.2018.1442239}
\showDOI{\tempurl}


\bibitem[Ishihara(1987)]%
        {Ishihara1987}
\bibfield{author}{\bibinfo{person}{Shinobu Ishihara}.}
  \bibinfo{year}{1987}\natexlab{}.
\newblock \bibinfo{booktitle}{\emph{Test for colour-blindness}}.
\newblock \bibinfo{publisher}{Kanehara Tokyo, Japan}.
\newblock


\bibitem[Izard(2010)]%
        {Izard2010}
\bibfield{author}{\bibinfo{person}{Carroll~E. Izard}.}
  \bibinfo{year}{2010}\natexlab{}.
\newblock \bibinfo{title}{The many meanings/aspects of emotion: Definitions,
  functions, activation, and regulation}.
\newblock , \bibinfo{numpages}{363-370}~pages.
\newblock
Issue 4.
\showISSN{17540739}
\urldef\tempurl%
\url{https://doi.org/10.1177/1754073910374661}
\showDOI{\tempurl}


\bibitem[Jansen and Fischbach(2019)]%
        {Jansen2019}
\bibfield{author}{\bibinfo{person}{Pascal Jansen} {and} \bibinfo{person}{Fabian
  Fischbach}.} \bibinfo{year}{2019}\natexlab{}.
\newblock \bibinfo{booktitle}{\emph{QuestionnaireToolkit for Unity}}.
\newblock
\urldef\tempurl%
\url{https://zefwih.com}
\showURL{%
\tempurl}


\bibitem[Jennett et~al\mbox{.}(2008)]%
        {Jennett2008}
\bibfield{author}{\bibinfo{person}{Charlene Jennett}, \bibinfo{person}{Anna~L.
  Cox}, \bibinfo{person}{Paul Cairns}, \bibinfo{person}{Samira Dhoparee},
  \bibinfo{person}{Andrew Epps}, \bibinfo{person}{Tim Tijs}, {and}
  \bibinfo{person}{Alison Walton}.} \bibinfo{year}{2008}\natexlab{}.
\newblock \showarticletitle{Measuring and defining the experience of immersion
  in games}.
\newblock \bibinfo{journal}{\emph{International Journal of Human-Computer
  Studies}} \bibinfo{volume}{66}, \bibinfo{number}{9} (\bibinfo{year}{2008}),
  \bibinfo{pages}{641--661}.
\newblock
\showISSN{1071-5819}
\urldef\tempurl%
\url{https://doi.org/10.1016/j.ijhcs.2008.04.004}
\showDOI{\tempurl}


\bibitem[Joosten et~al\mbox{.}(2012)]%
        {Joosten2012}
\bibfield{author}{\bibinfo{person}{Evi Joosten}, \bibinfo{person}{Giel~Van
  Lankveld}, {and} \bibinfo{person}{Pieter Spronck}.}
  \bibinfo{year}{2012}\natexlab{}.
\newblock \showarticletitle{Influencing Player Emotions Using Colors}.
\newblock \bibinfo{journal}{\emph{Journal of Intelligent Computing}}
  \bibinfo{volume}{3}, \bibinfo{number}{2} (\bibinfo{year}{2012}),
  \bibinfo{pages}{76--86}.
\newblock


\bibitem[Kim and Sundar(2014)]%
        {Kim2014}
\bibfield{author}{\bibinfo{person}{Ki~Joon Kim} {and} \bibinfo{person}{S.~Shyam
  Sundar}.} \bibinfo{year}{2014}\natexlab{}.
\newblock \showarticletitle{Does screen size matter for smartphones?
  Utilitarian and hedonic effects of screen size on smartphone adoption}.
\newblock \bibinfo{journal}{\emph{Cyberpsychology, Behavior, and Social
  Networking}}  \bibinfo{volume}{17} (\bibinfo{date}{7} \bibinfo{year}{2014}),
  \bibinfo{pages}{466--473}.
\newblock
Issue 7.
\showISSN{21522723}
\urldef\tempurl%
\url{https://doi.org/10.1089/cyber.2013.0492}
\showDOI{\tempurl}


\bibitem[Knez and Niedenthal(2008)]%
        {Knez2008}
\bibfield{author}{\bibinfo{person}{Igor Knez} {and} \bibinfo{person}{Simon
  Niedenthal}.} \bibinfo{year}{2008}\natexlab{}.
\newblock \showarticletitle{Lighting in digital game worlds: Effects on affect
  and play performance}.
\newblock \bibinfo{journal}{\emph{Cyberpsychology and Behavior}}
  \bibinfo{volume}{11} (\bibinfo{date}{4} \bibinfo{year}{2008}),
  \bibinfo{pages}{129--137}.
\newblock
Issue 2.
\showISSN{10949313}
\urldef\tempurl%
\url{https://doi.org/10.1089/cpb.2007.0006}
\showDOI{\tempurl}


\bibitem[Kowert and Quandt(2020)]%
        {Kowert2020}
\bibfield{author}{\bibinfo{person}{Rachel Kowert} {and}
  \bibinfo{person}{Thorsten Quandt}.} \bibinfo{year}{2020}\natexlab{}.
\newblock \bibinfo{booktitle}{\emph{The Video Game Debate 2}}.
\newblock \bibinfo{publisher}{Routledge}.
\newblock
\urldef\tempurl%
\url{https://doi.org/10.4324/9780429351815}
\showDOI{\tempurl}


\bibitem[Kuittinen et~al\mbox{.}(2007)]%
        {Kuittinen2007}
\bibfield{author}{\bibinfo{person}{Jussi. Kuittinen},
  \bibinfo{person}{Annakasia. Kultima}, \bibinfo{person}{Johannes. Niemelä},
  {and} \bibinfo{person}{Janne Paavalinen}.} \bibinfo{year}{2007}\natexlab{}.
\newblock \showarticletitle{Casual games discussion}.
\newblock \bibinfo{journal}{\emph{Proceedings of the 2007 Conference on Future
  Play.}} (\bibinfo{year}{2007}), \bibinfo{pages}{105--112}.
\newblock
\showISBNx{9781595939432}


\bibitem[Landers(1980)]%
        {Landers1980}
\bibfield{author}{\bibinfo{person}{Daniel~M. Landers}.}
  \bibinfo{year}{1980}\natexlab{}.
\newblock \showarticletitle{The Arousal-Performance Relationship Revisited}.
\newblock \bibinfo{journal}{\emph{Research Quarterly for Exercise and Sport}}
  \bibinfo{volume}{51}, \bibinfo{number}{1} (\bibinfo{year}{1980}),
  \bibinfo{pages}{77--90}.
\newblock
\urldef\tempurl%
\url{https://doi.org/10.1080/02701367.1980.10609276}
\showDOI{\tempurl}
\showeprint{https://doi.org/10.1080/02701367.1980.10609276}
\newblock
\shownote{PMID: 7394290}.


\bibitem[Loderer et~al\mbox{.}(2020)]%
        {Loderer2020}
\bibfield{author}{\bibinfo{person}{Kristina Loderer}, \bibinfo{person}{Reinhard
  Pekrun}, {and} \bibinfo{person}{James~C. Lester}.}
  \bibinfo{year}{2020}\natexlab{}.
\newblock \showarticletitle{Beyond cold technology: A systematic review and
  meta-analysis on emotions in technology-based learning environments}.
\newblock \bibinfo{journal}{\emph{Learning and Instruction}}
  \bibinfo{volume}{70} (\bibinfo{date}{12} \bibinfo{year}{2020}).
\newblock
\showISSN{09594752}
\urldef\tempurl%
\url{https://doi.org/10.1016/j.learninstruc.2018.08.002}
\showDOI{\tempurl}


\bibitem[Ouariachi et~al\mbox{.}(2018)]%
        {Ouariachi2018}
\bibfield{author}{\bibinfo{person}{Tania Ouariachi},
  \bibinfo{person}{Mar{\'i}a~Dolores Olvera-Lobo}, {and}
  \bibinfo{person}{Jos{\'e} Guti{\'e}rrez-P{\'e}rez}.}
  \bibinfo{year}{2018}\natexlab{}.
\newblock \bibinfo{booktitle}{\emph{Serious Games and Sustainability}}.
\newblock \bibinfo{publisher}{Springer}, \bibinfo{pages}{1--10}.
\newblock
\showISBNx{978-3-319-63951-2}
\urldef\tempurl%
\url{https://doi.org/10.1007/978-3-319-63951-2_326-1}
\showDOI{\tempurl}


\bibitem[Plass et~al\mbox{.}(2020)]%
        {Plass2020}
\bibfield{author}{\bibinfo{person}{Jan~L. Plass}, \bibinfo{person}{Bruce~D.
  Homer}, \bibinfo{person}{Andrew MacNamara}, \bibinfo{person}{Teresa Ober},
  \bibinfo{person}{Maya~C. Rose}, \bibinfo{person}{Shashank Pawar},
  \bibinfo{person}{Chris~M. Hovey}, {and} \bibinfo{person}{Alvaro Olsen}.}
  \bibinfo{year}{2020}\natexlab{}.
\newblock \showarticletitle{Emotional design for digital games for learning:
  The effect of expression, color, shape, and dimensionality on the affective
  quality of game characters}.
\newblock \bibinfo{journal}{\emph{Learning and Instruction}}
  \bibinfo{volume}{70} (\bibinfo{year}{2020}), \bibinfo{pages}{101194}.
\newblock
\showISSN{09594752}
\urldef\tempurl%
\url{https://doi.org/10.1016/j.learninstruc.2019.01.005}
\showDOI{\tempurl}


\bibitem[Plass and Kaplan(2016)]%
        {Plass2016}
\bibfield{author}{\bibinfo{person}{Jan~L. Plass} {and} \bibinfo{person}{Ulas
  Kaplan}.} \bibinfo{year}{2016}\natexlab{}.
\newblock \showarticletitle{Emotional Design in Digital Media for Learning}.
\newblock \bibinfo{publisher}{Elsevier}, \bibinfo{pages}{131--161}.
\newblock
\urldef\tempurl%
\url{https://doi.org/10.1016/b978-0-12-801856-9.00007-4}
\showDOI{\tempurl}


\bibitem[Posner et~al\mbox{.}(2005)]%
        {Posner2005}
\bibfield{author}{\bibinfo{person}{Jonathan Posner}, \bibinfo{person}{James~A
  Russell}, {and} \bibinfo{person}{Bradley~S Peterson}.}
  \bibinfo{year}{2005}\natexlab{}.
\newblock \showarticletitle{The circumplex model of affect: An integrative
  approach to affective neuroscience, cognitive development, and
  psychopathology}.
\newblock \bibinfo{journal}{\emph{Development and Psychopathology}}
  \bibinfo{volume}{17} (\bibinfo{year}{2005}), \bibinfo{pages}{715--734}.
\newblock
\urldef\tempurl%
\url{https://doi.org/10.10170S0954579405050340}
\showDOI{\tempurl}


\bibitem[Russell(1980)]%
        {Russell1980}
\bibfield{author}{\bibinfo{person}{James~A. Russell}.}
  \bibinfo{year}{1980}\natexlab{}.
\newblock \showarticletitle{A circumplex model of affect}.
\newblock \bibinfo{journal}{\emph{Journal of Personality and Social
  Psychology}} \bibinfo{volume}{39}, \bibinfo{number}{6}
  (\bibinfo{year}{1980}), \bibinfo{pages}{1161--1178}.
\newblock
\showISSN{00223514}
\urldef\tempurl%
\url{https://doi.org/10.1037/h0077714}
\showDOI{\tempurl}


\bibitem[Shashank'~'Tam(2019)]%
        {Pawar2019}
\bibfield{author}{\bibinfo{person}{Jan L.'~'Pawar Shashank'~'Tam,
  Frankie'~'Plass}.} \bibinfo{year}{2019}\natexlab{}.
\newblock \bibinfo{booktitle}{\emph{"Emerging design factors in game-based
  learning: Emotional design, musical score, and game mechanics design}}.
\newblock \bibinfo{publisher}{The MIT Press}, \bibinfo{pages}{347--366}.
\newblock


\bibitem[Tondello and Nacke(2019)]%
        {Tondello2019}
\bibfield{author}{\bibinfo{person}{Gustavo~F. Tondello} {and}
  \bibinfo{person}{Lennart~E. Nacke}.} \bibinfo{year}{2019}\natexlab{}.
\newblock \showarticletitle{Player characteristics and video game preferences}.
\newblock \bibinfo{journal}{\emph{CHI PLAY 2019 - Proceedings of the Annual
  Symposium on Computer-Human Interaction in Play}}, \bibinfo{pages}{365--378}.
\newblock
\showISBNx{9781450366885}
\urldef\tempurl%
\url{https://doi.org/10.1145/3311350.3347185}
\showDOI{\tempurl}


\bibitem[Vallat(2018)]%
        {Vallat2018}
\bibfield{author}{\bibinfo{person}{Raphael Vallat}.}
  \bibinfo{year}{2018}\natexlab{}.
\newblock \showarticletitle{Pingouin: statistics in Python}.
\newblock \bibinfo{journal}{\emph{Journal of Open Source Software}}
  \bibinfo{volume}{3}, \bibinfo{number}{31} (\bibinfo{year}{2018}),
  \bibinfo{pages}{1026}.
\newblock
\urldef\tempurl%
\url{https://doi.org/10.21105/joss.01026}
\showDOI{\tempurl}


\bibitem[Wilms and Oberfeld(2018)]%
        {Wilms2018}
\bibfield{author}{\bibinfo{person}{Lisa Wilms} {and} \bibinfo{person}{Daniel
  Oberfeld}.} \bibinfo{year}{2018}\natexlab{}.
\newblock \showarticletitle{Color and emotion: effects of hue, saturation, and
  brightness}.
\newblock \bibinfo{journal}{\emph{Psychological Research}}
  \bibinfo{volume}{82}, \bibinfo{number}{5} (\bibinfo{year}{2018}),
  \bibinfo{pages}{896--914}.
\newblock
\showISSN{14302772}
\urldef\tempurl%
\url{https://doi.org/10.1007/s00426-017-0880-8}
\showDOI{\tempurl}


\end{thebibliography}

\appendix

\end{document}